\documentclass[
reprint,
superscriptaddress,
 amsmath,amssymb,
aip,
pop,
floatfix,
]{revtex4-1}

\usepackage{graphicx}
\usepackage{epstopdf}
\usepackage{dcolumn}
\usepackage{bm}
\usepackage{ifpdf}
\usepackage[usenames,dvipsnames]{color}
\usepackage{hyperref}
\def\Snospace~{\S{}}

\usepackage[all]{hypcap}
\hypersetup{urlcolor=Blue, citecolor=Blue,linkcolor=Blue,colorlinks=true}

\newcommand{\sref}[2]{\hyperref[#1]{Fig. \ref{#1}#2}}

\newcommand{\Fig}[1]{
	\label{fig:#1}
	\ifpdf
	\includegraphics[scale=1]{apsdpp2017_figures_#1.png} 
	\else
	\includegraphics[scale=1]{apsdpp2017_figures_#1.eps}
	\fi
}

\newcommand{\x}[1]{$x=#1$ mm}
\newcommand{\y}[1]{$y=#1$ mm}

\renewcommand{\t}[1]{$t=#1$ ns}

\newcommand{\xcmcubed}[2]{$#1\times10^{#2}$ cm$^{-3}$}
\renewcommand{\deg}[1]{#1$^{\circ}$}

\begin{document}

\title{An Experimental Platform for Pulsed-Power Driven Magnetic Reconnection}

\author{J. D. Hare}
\email{jdhare@imperial.ac.uk}
\author{L. G. Suttle}
\author{S. V. Lebedev}
\email{s.lebedev@imperial.ac.uk}
\affiliation{Blackett Laboratory, Imperial College, London, SW7 2AZ, United Kingdom}
\author{N. F. Loureiro}
\affiliation{Plasma Science and Fusion Center, Massachusetts Institute of Technology, Cambridge MA 02139, USA}
\author{A. Ciardi}
\affiliation{Sorbonne Universit\'{e}s, UPMC Univ Paris 06, Observatoire de Paris, PSL Research University, CNRS, UMR 8112, LERMA, F-75005, Paris, France}
\author{J. P. Chittenden}
\author{T. Clayson}
\author{S. J. Eardley}
\author{C. Garcia}
\author{J. W. D. Halliday}
\author{T. Robinson}
\author{R. A. Smith}
\author{N. Stuart}
\author{F. Suzuki-Vidal}
\author{E. R. Tubman}
\affiliation{Blackett Laboratory, Imperial College, London, SW7 2AZ, United Kingdom}

\date{\today}

\begin{abstract}
We describe a versatile pulsed-power driven platform for magnetic reconnection experiments, based on exploding wire arrays driven in parallel [Suttle, L. G. \textit{et al.} PRL, 116, 225001].
This platform produces inherently magnetised plasma flows for the duration of the generator current pulse (250 ns), resulting in a long-lasting reconnection layer.
The layer exists for long enough to allow evolution of complex processes such as plasmoid formation and movement to be diagnosed by a suite of high spatial and temporal resolution laser-based diagnostics.
We can access a wide range of magnetic reconnection regimes by changing the wire material or moving the electrodes inside the wire arrays.
We present results with aluminium and carbon wires, in which the parameters of the inflows and the layer which forms are significantly different.
By moving the electrodes inside the wire arrays, we change how strongly the inflows are driven.
This enables us to study both symmetric reconnection in a range of different regimes, and asymmetric reconnection.
\end{abstract}

\maketitle
\section{Introduction}

Magnetic reconnection is a fundamental phenomenon in plasma physics\cite{Yamada2010,Zweibel2009,Zweibel2016}, and occurs across a vast range of parameter space, from the dense, collisional plasmas inside the solar convective zone\cite{Ryutov2015a} to the radiatively cooled plasmas surrounding pulsars\cite{Uzdensky2011} or the asymmetrically driven reconnection events in the Earth's magnetosphere.

Studying this range of plasma conditions in the laboratory is an ongoing challenge, with experiments such as MRX\cite{Ji1999,Yamada2015a,Yamada2015}, TREX\cite{Olson2016} and LAPD\cite{Gekelman2016} producing results in externally magnetised gas discharge plasmas, as well as experiments studying reconnection in plasmas produced by laser beams, with self-generated\cite{Nilson2006, Willingale2010} or externally imposed\cite{Fiksel2014b} magnetic fields.
Any given experiment is limited in the range of plasma conditions it can achieve, which motivates the development of new experimental platforms which can provide complementary measurements.
These enable us to determine the generic, rather than experiment-specific, features of magnetic reconnection.

One outstanding question in magnetic reconnection is the stability of the reconnection layer.
Recent theoretical work\cite{Loureiro2007a, Loureiro2015} has shown that the reconnection layer can break up into a series of magnetic islands, known as plasmoids.
These plasmoids dramatically increase the rate at which magnetic flux can be annihilated by the reconnecting current sheet,\cite{Bhattacharjee2009,Huang2011,Uzdensky2010,Loureiro2012} and may be responsible for enhanced heating of the electrons and ions\cite{Loureiro2012}.

In the collisional regime, where magneto-hydro-dynamics (MHD) is a good description of the plasma, a high Lundquist number $S=\mu_0 L V_A/\eta>S_C\sim10^4$ (where $L$ is the half-length of the current sheet, $V_A$ is the Alfv\'en velocity and $\eta$ is the Spitzer-Braginskii plasma resistivity) is required for plasmoids to form (see Ref. \onlinecite{Loureiro2013} and references to numerical simulations therein).
The transition to a plasmoid unstable regime at a critical Lundquist number has been observed in simulations, but has not yet been experimentally verified in the collisional regime.
However, there are other regimes, such as the semi-collisional regime\cite{Baalrud2011, Loureiro2015}, in which it is now understood that the boundaries do not include a lower bound on the Lundquist number.\cite{Loureiro2015}
This allows plasmoids to form under conditions more easily accessed by experiments and simulations.
The diversity of plasmas in which plasmoids can form is vital for understanding their role in magnetic reconnection throughout the universe,\cite{Ji2011} and our appreciation of this diversity can be improved through dedicated laboratory experiments.\cite{Olson2016,Jara-almonte2016,Hare2017,Hare2017c}

In this paper we present results from the recently developed\cite{Suttle2016,Hare2017,Hare2017c, Hare2017a} pulsed-power-driven magnetic reconnection platform at the \textsc{Magpie} generator at Imperial College London.\cite{Mitchell1996}
This platform offers a versatile and highly configurable set up which enables the parameters of the inflowing plasma to be altered by changing the material of the plasma or the geometry of the exploding wire arrays.

This paper is organised as follows: we begin in \autoref{sec:experimental_setup} by discussing the driver and plasma source for these experiments, as well as the reconnection geometry used in this paper. 
In \autoref{sec:wire_material} we discuss experiments in which we used aluminium or carbon wires, and in \autoref{sec:differences} we focus on the similarities and differences that we observe in these two cases.
In \autoref{sec:geometry} we present a modification to the existing experimental set-up, in which we move the thick electrodes inside the wire arrays to change the driving force on the wires, and we present results which show marked differences in the layer electron density.
In \autoref{sec:asymmetric} we further modify this set-up to produce asymmetric inflows, and we observe that the reconnection layer is significantly displaced from the mid-plane.
We conclude and present our outlook for future experiments in \autoref{sec:conclusions}.

\section{Experimental Setup}\label{sec:experimental_setup}
\begin{figure*}[t]
	\Fig{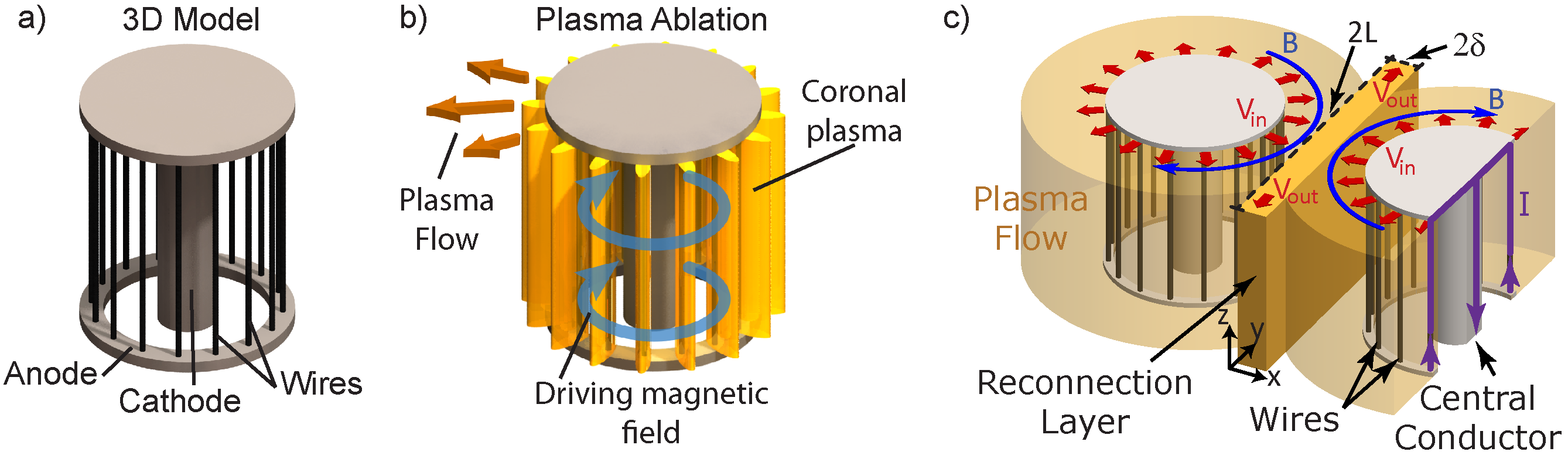}
	\centering
	\caption{The plasma source for these magnetic reconnection experiments, the exploding (``inverse'') cylindrical wire array. a) The bare array, before the current pulse, consisting of a cylindrical cage of thin wires surrounding a thick central conductor. b) The array during the current pulse, with a coronal plasma caused by Ohmic heating of the wire surfaces which is accelerated outwards by the large azimuthal magnetic field surrounding the central conductor. c) Two exploding wire arrays driven in parallel. The flows are inherently magnetised and flow radially outwards from each array, colliding at the mid-plane with anti-parallel magnetic fields which undergo magnetic reconnection.}
\end{figure*}

\subsection{Load Hardware}

The load hardware consisted of two cylindrical exploding (``inverse'') wire arrays\cite{Harvey-Thompson2009}, driven in parallel by the \textsc{Magpie} pulsed power generator (1.4 MA peak current, 240 ns rise time).
Each array was 16 mm in diameter, 16 mm tall and consisted of 16 equally space wires surrounding the central conductor (5 mm diameter, stainless steel) (\sref{fig:setup}{a}).
In this paper, we either used 30 $\mu$m diameter aluminium wires (California Fine Wire Company) or $300\,\mu$m diameter carbon wires (Staedler Mars Micro Carbon B).

Initially the drive current heats the surface of the wires, producing a cold coronal plasma\cite{Lebedev2001}.
The magnetic field configuration at each wire is dominated by the contribution from the central conductor, and a fraction of the drive current switches into the plasma.
The coronal plasma therefore experiences a strong $\textbf{J}\times\textbf{B}$ radially outwards, which accelerates the plasma away from the wires (\sref{fig:setup}{b}).
As the plasma travels outwards, it advects a fraction of the driving azimuthal magnetic field, and so the flows from the two exploding wire arrays are inherently magnetised.
The coronal plasma is continually replenished and accelerated outwards for the duration of the \textsc{Magpie} current pulse ($\sim$ 500 ns), delivering a long lasting source of magnetised, super-sonic plasma.

In order to study magnetic reconnection, two of these arrays are placed side by side and driven in parallel  (\sref{fig:setup}{c}).
The two arrays produce radially diverging plasma flows which advect azimuthal magnetic fields.
When the flows collide, the magnetic fields they carry are anti-parallel and magnetic reconnection occurs with the formation of a long-lasting current sheet ($>200$ ns) in which the magnetic field is annihilated.

\subsection{Diagnostic Setup}
\begin{figure*}[t]
	\Fig{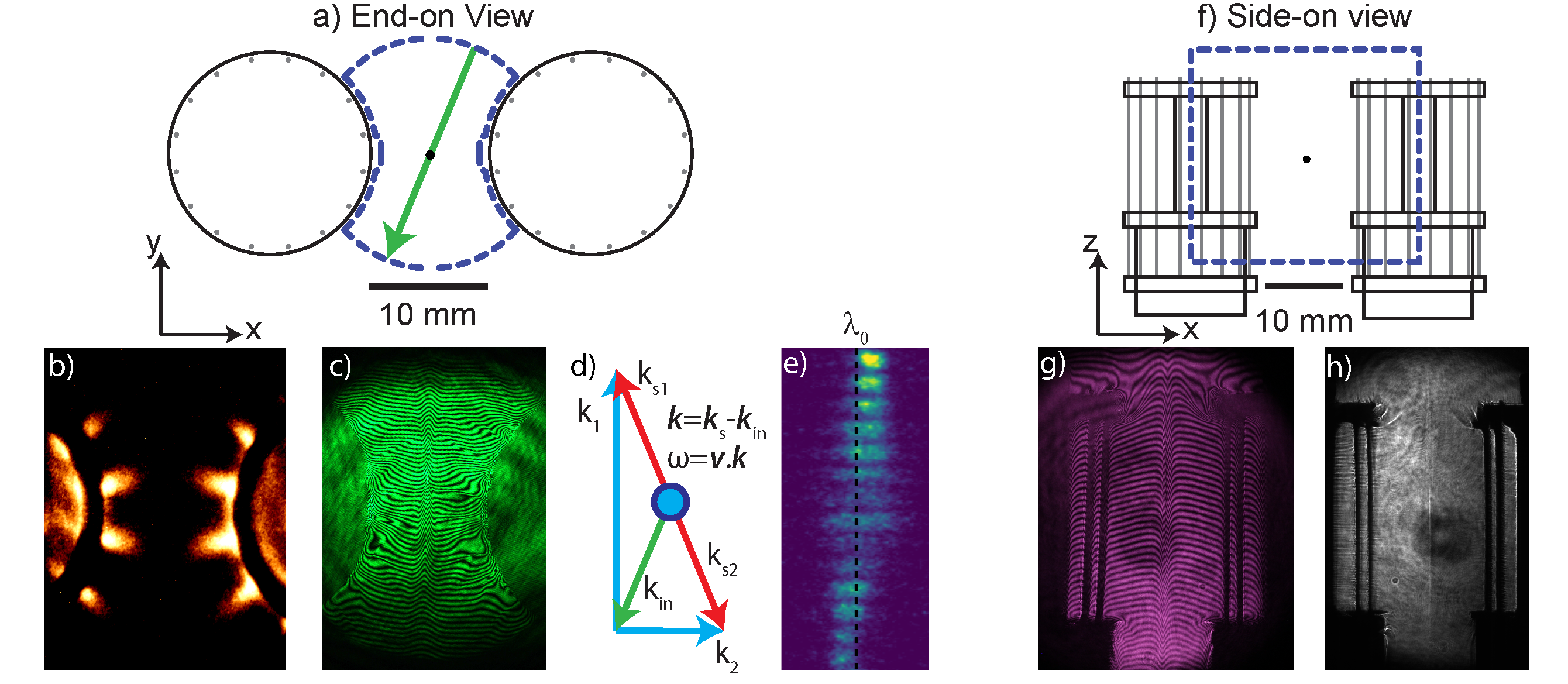}
	\centering
	\caption{The sight lines used by diagnostics in these experiments, along with example data from the diagnostics. a) End-on view, looking down onto the reconnection ($x,y$) plane. Field-of-view shown as a dashed blue line, dot marking the origin of the coordinate system. Path of Thomson scattering beam shown as a green arrow. b) Optical self emission image captured by a fast-framing camera (5 ns exposure). c) Laser interferometry for end-on electron density measurements. d) Scattering vector diagram for Thomson scattering, showing the two scattering vectors $k_{s1}$ and $k_{s2}$ corresponding to the two fibre optic bundles, and the resultant vectors each bundle is sensitive to Doppler shifts along. e) Sample data from Thomson scattering. f) Side-on view, looking at the ($x,z$) plane in which the reconnecting magnetic fields points into and out of the page. g) Laser interferometry for side-on electron density measurements. h) Image from one channel of the polarimetry diagnostic, showing increased intensity on the left and decreased intensity on the right of the image due to a change in polarisation state.}
\end{figure*}

The open nature of the load geometry allows for a suite of high resolution, spatially and temporally resolved diagnostics to study the reconnection layer along multiple lines-of-sight.

An overview of the reconnection dynamics was provided by an optical fast-framing camera.
This camera was sensitive to the optical self emission from the plasma at wavelengths $>600$ nm (a low-pass filter was used to block laser light), and is capable of taking 12 images with an inter-frame and exposure time as low as 5 ns.

Mach-Zehnder laser imaging interferometry was used to measure the line integrated electron density in both the end on ($x-y$ plane) and side-on ($z-x$ plane) directions. 
End-on interferometry was carried out using the 2nd (532 nm) and 3rd (355 nm) harmonics of a Nd-YAG laser (EKSPLA SL321P, 500 ps, 500 mJ)\cite{Swadling2014a}, with the beams imaged onto Canon 350D and 500D cameras.
The cameras were triggered with a long exposure (1.3 s), and so the time resolution of the images was set by the laser pulse length (500 ps).
The shift in the interference fringes is proportional to the line integrated electron density, $F\propto\int n_e dl$, and the volumetric electron density $n_e$ can be recovered by dividing by a representative length scale of the plasma, given sufficient uniformity along the probing direction.
In these experiments, the plasma was uniform over a distance $h=16$ mm in the $z$ direction, and so for end-on imaging we calculated $n_e(x,y)=\int n_e(x,y,z) dz/h$.

The side-on interferometry was performed using an infra-red laser beam (1053 nm, 5 J, 1 ns)\cite{Swadling2014a}, imaged onto a Canon 500D with a long exposure (1.3 s). 
This laser beam was simultaneously used for polarimetry measurements, in which the angle of the linear polarisation of the probing beam is determined by two CCDs (ATIK 383L+), behind high contrast, wide acceptance angle polarisers at equal and opposite angles (usually \deg{3}) to the extinction angle of the initial polarisation.
These allowed the change in linear polarisation ($\alpha$) due to the Faraday effect, $\alpha(x,z)\propto\int n_e (x,y,z) \textbf{B}(x,y,z)\cdot d\textbf{l}$, to be measured.
An estimate for the average magnetic field aligned with the probing laser ($B_y$) can be found by dividing the rotation angle by $\int n_e dy$ from the in-line interferometer, giving $B_y (x,z)\propto\alpha(x,z)/\int n_e(x,y,z) dy$.
This procedure is discussed in more detail in Ref. \onlinecite{Swadling2014a}.

Thomson scattering was used to measure the flow velocity and electron and ion temperatures in the inflows and the reconnection layer.
A focused laser beam (532 nm, 2 J, 8 ns) passed through the plasma, and the scattered light was collected by fibre optic bundles.
Each fibre optic bundle consisted of fourteen fibre optics, and so light was collected for fourteen distinct spatial locations.
The ion feature dominated the Thomson scattering spectra, and this enabled the electron and ion temperatures to be determined by fitting the experimental spectra with a theoretical model.\cite{Froula2011,Hare2017a}
Thomson scattering determines $\bar{Z}T_e$ and this was factored into $\bar{Z}$ and $T_e$ using a non-Local-Thermodynamic-Equilibrium (nLTE) code.\cite{Chittenden2016}

One common configuration of the Thomson scattering diagnostics was to place the two fibre optics bundles on opposite sides of the vacuum chamber, at \deg{45} and \deg{135} to the probing laser, which passed at \deg{22.5} to the reconnection layer (the $y$ axis) [as in Ref. \onlinecite{Swadling2014}]. 
This configuration meant that the two bundles were sensitive to orthogonal velocity components aligned with the inflow ($V_x$) and outflow ($V_y$) velocities respectively.

\section{Results from Varying the Wire Material}
\label{sec:wire_material}
The experimental set-up described in the previous section can be used with a range of wire materials.
The choice of wire material has a profound effect on the physics of the reconnection processes, significantly altering the nature of the inflows and the stability of the reconnection layer.
This platform has been used to study reconnection with wire materials including aluminium, carbon, copper, stainless steel and tungsten.
In this paper we present a comparison of results from aluminium and carbon wires, for which the most detailed experiments and analysis have been carried out.
In these experiments, the two arrays were placed with a centre to centre distance of 27 mm, giving a radius of curvature of the magnetic field lines as $R_C= 13.5$ mm.
This sets the layer half-length $L=7$ mm, which is used as a characteristic length scale in the calculations below.

The key differences in the inflow parameters between aluminium and carbon are the Mach numbers --- the flows are super-Alfv\'enic for aluminium and sub-Alfv\'enic for carbon --- and the radiative cooling time-scale, which is shorter in aluminium than in carbon, resulting in colder inflows.

\subsection{Results from Experiments with Aluminium Wires}

\begin{figure*}[t]
	\Fig{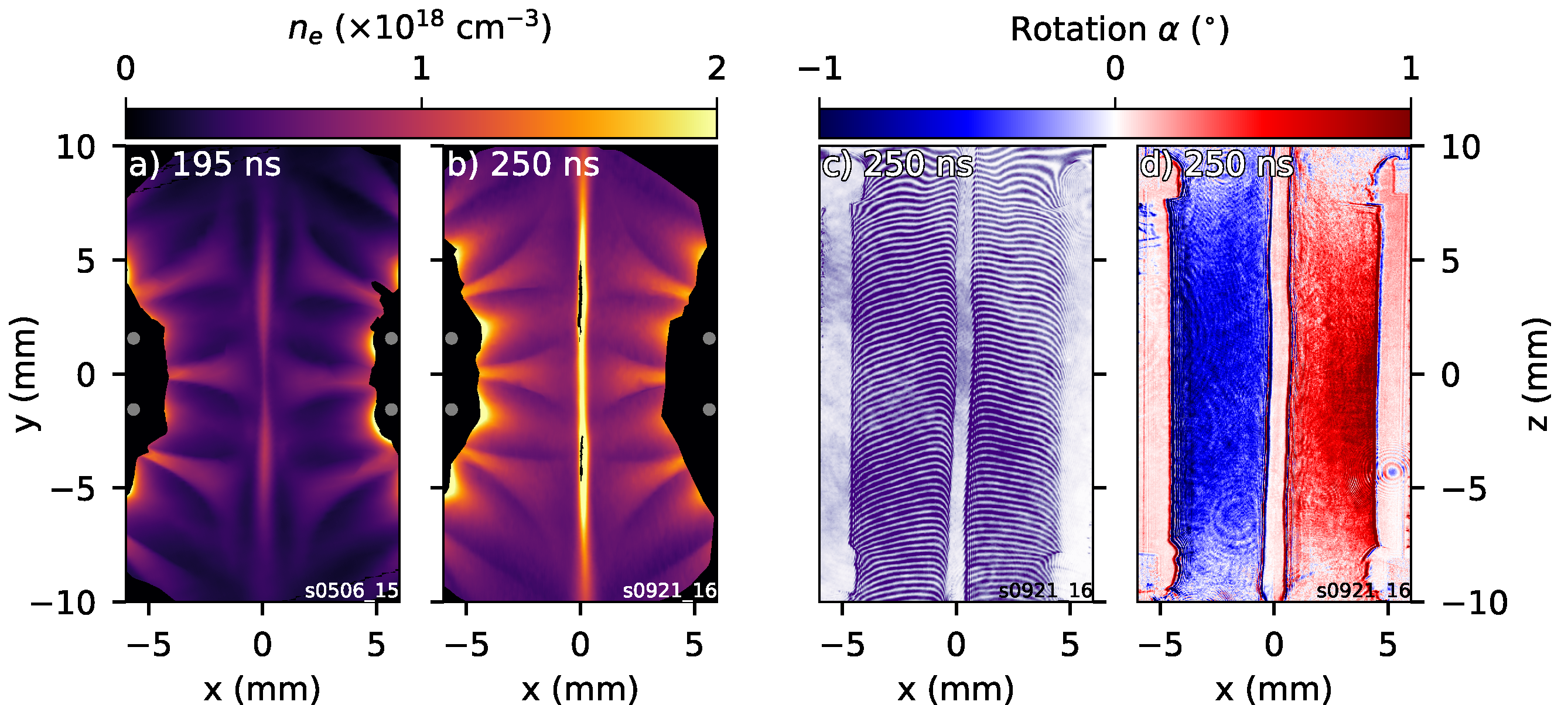}
	\centering
	\caption{Results from laser interferometry and polarimetry with aluminium wires. a) End-on electron density map at \t{195} after current start, shortly after the flows initially collide, wire locations marked in grey. b) End-on electron density map at \t{250} after current start, when the ion temperature has rapidly dropped and the electron density has increased. The interferogram becomes unreadable in the black regions, and so not electron density data is available there. c) Side-on interferogram showing increase in density as the flows collide at the mid-plane. d) Polarimetry map from \t{250} after current start [the same shot as in b)], showing the sharp features on either side of the reconnection layer which imply a jump in the magnetic field.}
\end{figure*}

Aluminium wires produced super-sonic ($M_S=3.1$), super-Alfv\'enic ($M_A=2.2$) inflows, which advected a $B=2$T magnetic field at a flow velocity of $V_{in}=50$ km/s.\cite{Suttle2016}
These inflows formed as the result of oblique shocks between the ablation streams from adjacent wires, which redirected the flows and formed a dense, collimated jet which propagated radially outwards from the gap between adjacent wires (\sref{fig:al_ne_alpha}{a}).\cite{Lebedev2014}
Although the electron density in the inflows is highly modulated, the oppositely directed flows from the two wire arrays collide at the mid-plane to form a well defined, uniform reconnection layer which persists for over 200 ns.

There is a substantial increase in the electron density inside the layer, consistent with the presence of a standing shock.
The reconnection layer exhibits two peaks in the electron density along \x{0}, which form symmetrically on either side of \y{0} and propagate outwards at around $V_y$=30--50 km/s.
The electron density can also be seen in \sref{fig:al_ne_alpha}{c}, which shows the side-on interferogram.
Here the interference fringes rise sharply at the boundary with the reconnection layer.
In the centre of the layer the electron density gradients are so large that the probing laser beam is deflected out of the collection cone of the imaging optics, and so here the interference fringes are lost.

The inflows advect magnetic field, as seen in polarisation angle map produced by Faraday rotation imaging (\sref{fig:al_ne_alpha}{d}).
The field in the inflowing plasma was around $B=2$T, and the field was annihilated within a narrow region around $2\delta\approx 0.6$ mm wide.
Sharp features are visible in the magnetic field maps on either side of the reconnection layer, which suggests that the magnetic field piles up outside the layer (at \x{\pm5}), consistent with the super-Alfv\'enic nature of the inflows (see Ref. \onlinecite{Suttle2016}, Fig. 5 for more details).

\begin{figure*}[t]
	\Fig{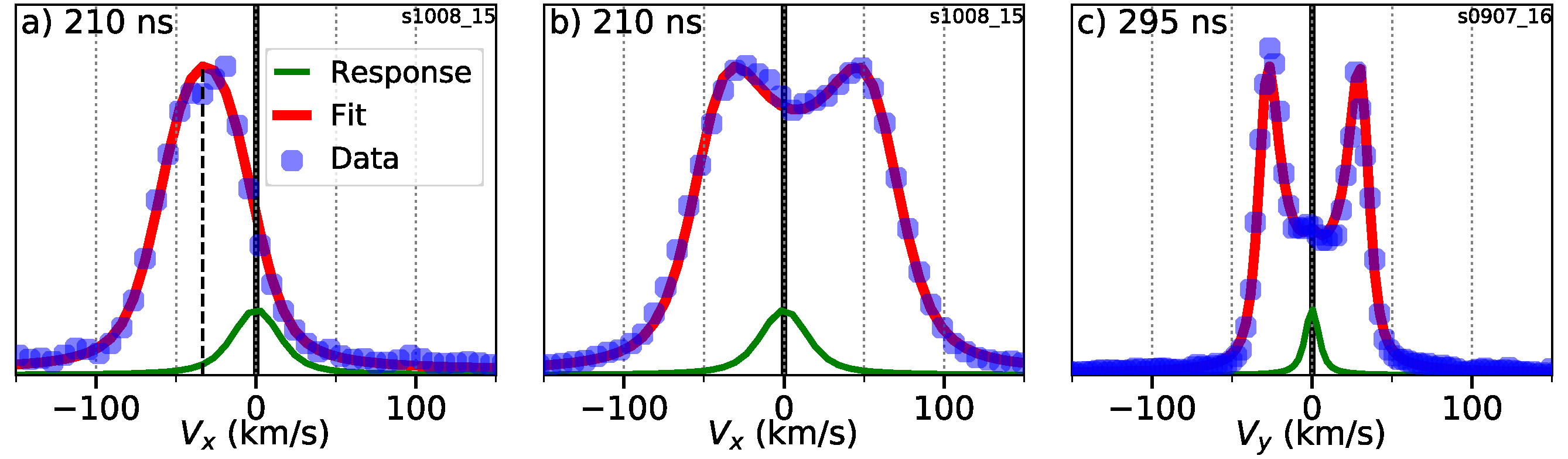}
	\centering
	\caption{Thomson Scattering results from experiments with aluminium wires. a) Spectrum of scattered light from a volume in the inflowing plasma, showing a Doppler shift to shorter wavelengths. b) Spectrum of scattered light from the same shot, from plasma at the centre of the reconnection layer, showing Doppler broadening and the presence of ion-acoustic peaks. c) Spectrum of scattered light from a different shot, at the centre of the reconnection layer at a later time when the ions have cooled, causing the ion-acoustic peaks to be more pronounced due to decreased thermal broadening. The narrower response function is due to converting between wavelength and velocity.}
\end{figure*}

Thomson scattering measurements show that the inflows approach the layer at $V_{in}\approx 50$ km/s, with an electron and ion temperature of $T_e\approx T_i\approx $ 25 eV (\sref{fig:al_thomson}{a}).
Inside the reconnection layer, the ion temperature rises to $T_i\approx $ 270 eV (\sref{fig:al_thomson}{b}). until around 240 ns, when it suddenly reduces to $T_i\approx $ 30 eV (\sref{fig:al_thomson}{c}).
The electron temperature rises to $T_e\approx $ 40 eV before \t{240}, before decreasing slightly to $T_e\approx $ 30 eV.
This rapid decrease in ion thermal energy is accompanied by a large increase in electron density, which increases from $n_e=$ \xcmcubed{1.3}{18} to $n_e=$ \xcmcubed{2.3}{18}.

Further measurements with Thomson scattering showed that the plasma is accelerated out of the layer (in the $y$ direction), reaching 80 km/s at \y{6}.

\subsection{Results from Experiments with Carbon Wires}

\begin{figure*}[t]
	\Fig{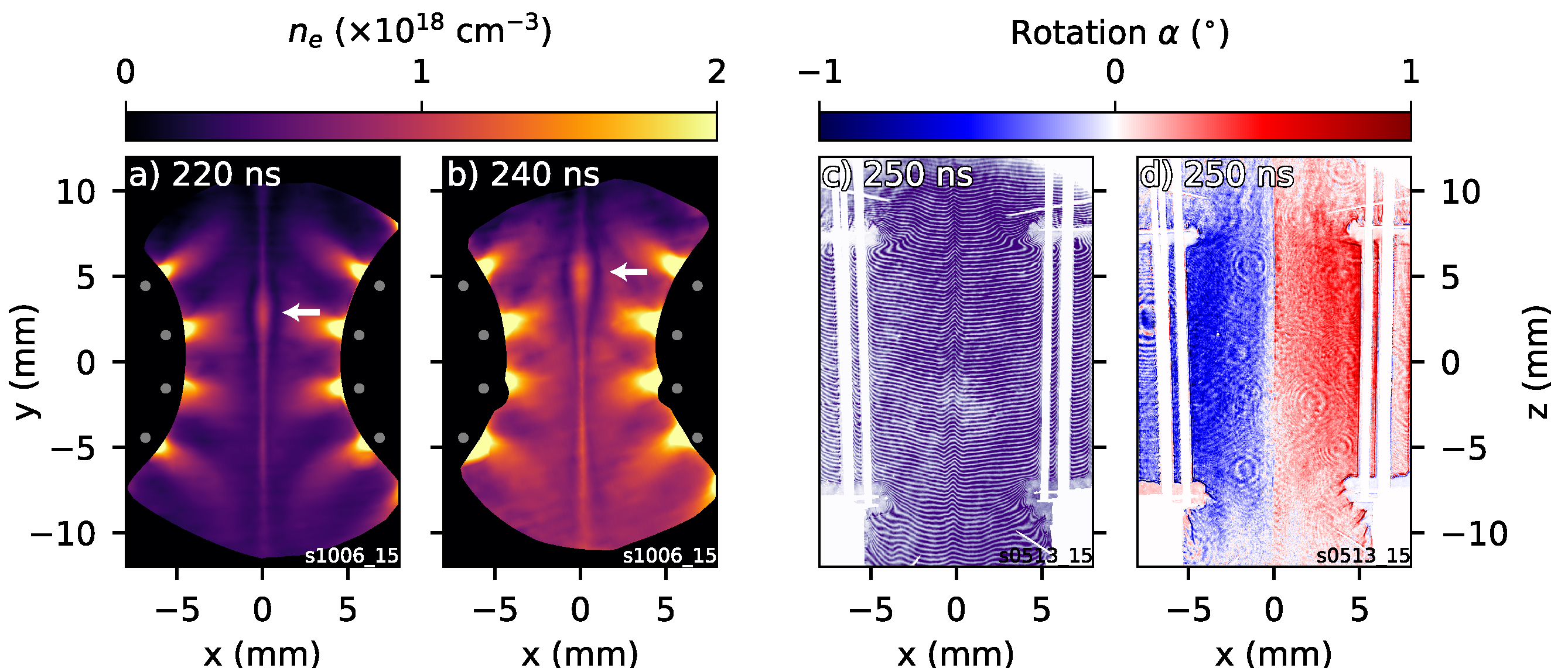}
	\centering
	\caption{Results from laser interferometry and polarimetry with carbon wires. a) End-on electron density map at \t{220} after current start, showing a reconnection layer which extends across the entire field of view, with a prominent plasmoid. Wire locations marked in grey. b) End-on electron density map at \t{240} after current start, from the same shot as a). Here the plasmoid has moved in the $+y$ direction at an apparent velocity of $V_y=130$ km/s. c) Polarimetry map from \t{250} after current start, showing the left/right anti-symmetric inflows with a smooth region in the centre in which the magnetic field is annihilated.}
\end{figure*}

Experiments with carbon wires produced super-sonic ($M_S=1.7$) but sub-Alfv\'enic ($M_A=0.7$) flows, which were accelerated to $V=50$ km/s and advected a field of $B=3$ T.\cite{Hare2017,Hare2017c}
The flows propagated radially outwards from each wire, without shock formation between adjacent flows (\sref{fig:c_ne_alpha}{a}), which suggests the flows have a fast magneto-sonic Mach number [$M_{FMS}^2=M_S^2 M_A^2 /(M_S^2+ M_A^2)$] less than unity.
These flows merged before reaching the mid-plane, and so formed a smooth inflow density profile to the reconnection layer, as discussed in Ref. \onlinecite{Hare2017c}.

The reconnection layer was observed to exist for over 200 ns, but although its width was constant over this time scale, the layer was not uniform or stable.
Instead, the layer was observed the rapidly break up into a chain of elliptical density perturbations, known as ``plasmoids''.
\footnote{We note that plasmoids have also been observed in experiments with 32 (rather than 16) wires per array.\cite{Hare2017a}}
These plasmoids were often observed in both of the two frames of interferometry, as shown in \sref{fig:c_ne_alpha}{a and b}.
Here the plasmoid moved 2.5 mm in 20 ns, which corresponds to an outflow velocity of around 130 km/s, which is consistent with the outflow velocity measured via Thomson scattering.
Magnetic probe measurements (discussed in Ref. \onlinecite{Hare2017c}) have shown that these plasmoids have the O-point magnetic field structure predicted by tearing instability theory.

The side-on electron density map (\sref{fig:c_ne_alpha}{c}) shows that the layer is uniform in the out-of-plane ($z$) direction, without the strong jumps in electron density associated with the layer formed by aluminium wires. 
The lack of shocks suggests that the magnetic field should not pile-up, which is confirmed by the polarogram in (\sref{fig:c_ne_alpha}{d}), which shows a smooth variation of rotation angle from  $ \pm$\deg{0.5} to \deg{0} within 1 mm.

The magnetic field map is produced by combining the electron density map, determined from \sref{fig:c_ne_alpha}{c}, and the polarogram shown in (\sref{fig:c_ne_alpha}{d}).
Lineouts from this map show a magnetic field profile which is well approximated by the Harris sheet, $B=B_0 \tanh(x/\delta)$, with $B_0= 3$ T and $\delta=0.6$ mm, as seen in Ref. \onlinecite{Hare2017c} Fig. 9.

\begin{figure*}[t]
	\Fig{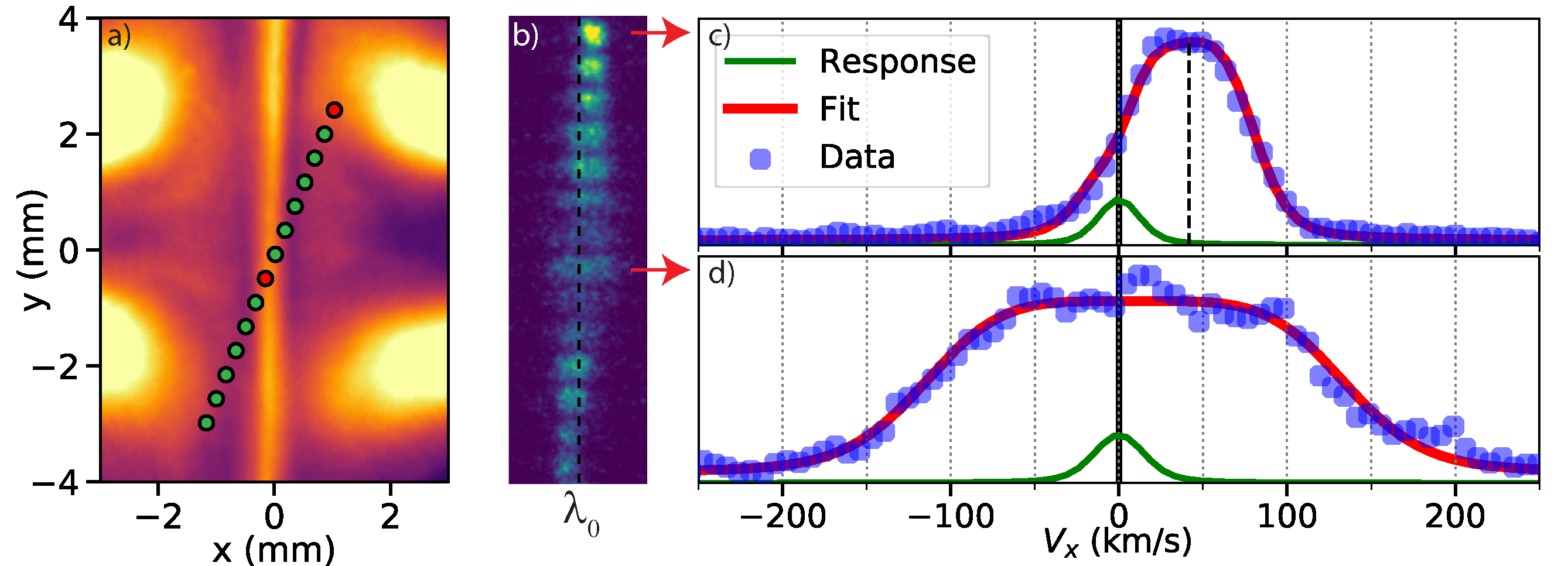}
	\centering
	\caption{Thomson Scattering results from experiments with carbon wires. a) Section of an electron density map with the collection volumes plotted as circles. The two red circles correspond to the spectra shown in c) and d). b) Raw spectrogram showing the spectra of light scattered at \deg{45} to the probing laser beam, which carries information about velocities in the inflow ($x$) direction. c) Sample spectrum from a collection volume in the inflowing plasma, outside the reconnection layer, showing a significant Doppler shift to longer wavelengths, and some Doppler broadening. d) Sample spectrum from a collection volume inside the reconnection layer showing no overall Doppler shift, but significant Doppler broadening.}
\end{figure*}

The reconnection layer was studied in more detail using the Thomson scattering diagnostic.
The focused laser beam passed through the chamber at \deg{22.5} to the reconnection layer, and scattered light was collected from 14 distinct spatial locations, as shown in \sref{fig:c_thomson}{a}.
The raw spectrogram from one of the fibre optic bundles is shown in \sref{fig:c_thomson}{b}, with the initial laser wavelength marked with a vertical dashed black line.

Examination of the raw spectrogram shows that for collection volumes outside the reconnection layer, the Doppler shift is significant and the Doppler broadening relatively small, and that the Doppler broadening increases towards \x{0} as the Doppler shift decreases.
This is borne out by the two sample spectra in \sref{fig:c_thomson}{c and d}.
The first spectrum (\sref{fig:c_thomson}{c}) comes from a collection volume which is furthest away from the centre of the reconnection layer, where the Doppler shift of +0.6 \AA{}  corresponds to an inflow velocity of $V_x =- 50$ km/s, with the negative sign due to the plasma flowing from the right array to the mid-plane.

The second spectrum (\sref{fig:c_thomson}{d}) was collected from the centre of the reconnection layer, where there is significant Doppler broadening. 
The flat-top nature of the spectrum implies that $\bar{Z}T_e\approx T_i$.

\section{Discussion of Results from Varying Wire Material}\label{sec:differences}

\begin{table*}[t]
\centering
\caption{Summary of parameters for experiments with aluminium and carbon wires. For aluminium, the arrow ($\rightarrow$) represents the change in parameters between \t{190} and \t{250}.}
\label{tab:summary}
\noindent\hrulefill 

\begin{tabular*}{\textwidth}{ @{\extracolsep{\fill}} lccccc}
                                &                 & \multicolumn{2}{c}{Aluminium}                         & \multicolumn{2}{c}{Carbon}            \\
Parameter                       &                 & Flow              & Layer                             & Flow              & Layer             \\\hline\hline\rule{0pt}{2.5ex}
Electron density, cm$^{-3}$     & $n_e$           & $5\times10^{17}$  & $1.4\rightarrow 2.3\times10^{18}$ & $3\times10^{17}$  & $6\times10^{17}$  \\
Effective charge                & $\bar{Z}$       & 3.5               & 7$\rightarrow$5.7                 & 4                 & 6                 \\
Electron temperature, eV        & $T_e$           & 15                & 40$\rightarrow$30                 & 15                & 100               \\
Ion temperature, eV             & $T_i$           & 22                & 270$\rightarrow$30                & 50                & 600               \\
Magnetic field, T               & $B_y$           & 2$\rightarrow$4   & --                                & 3                 & --                \\\hline
&                 &                   &                                   &                   &                   \\
Layer half-length, mm           & $L=R_C/2$       & --                & 7                                 & --                & 7                 \\
Layer half-thickness, mm        & $\delta$        & --                & 0.3                               & --               & 0.6               \\
Ion skin depth, mm              & $c/\omega_{pi}$ & 0.89              & 0.37$\rightarrow$0.33             & 0.71              & 0.41              \\
Ion-ion mean free path, mm      & $\lambda_{i,i}$ & 6$\times 10^{-4}$ & 3$\times 10^{-3}$                 & 4$\times 10^{-2}$ & 3$\times 10^{-3}$ \\\hline
&                 &                   &                                   &                   &                   \\
Inflow (outflow) velocity, km/s & $V_x$ ($V_y$)   & 50                & (100)                             & 50                & (130)             \\
Alfv\'en speed, km/s            & $V_A$           & 22$\rightarrow$35 & --                                & 70                & --                \\
Sound speed, km/s               & $C_S$           & 16                & 44$\rightarrow$27                 & 30                & 85                \\
Fast-magnetosonic speed, km/s            & $V_{FMS}$           & 24$\rightarrow$39 & --                                & 75                & --                \\\hline
&                 &                   &                                   &                   &                   \\
Ion-electron cooling time, ns   & $\tau^E_{e/i}$  & 50                & 40$\rightarrow$20                 & 30                & 140               \\
Radiative cooling time, ns      & $\tau_{rad}$    & 20                & 5$\rightarrow$3                   & 100               & 600               \\\hline
&                 &                   &                                   &                   &                   \\
Thermal beta                    & $\beta_{th}$    & 1.1               & --                                & 0.4               & --                \\
Dynamic beta                    & $\beta_{dyn}$   & 10                & --                                & 1                 & --                \\
Lundquist number                & $S$             & --                & 11$\rightarrow$ 7                 & --                & 120               \\
Two-fluid effects               & $L/d_i$         & --                & 19$\rightarrow$22                 & --                & 18    \\      
\hline\hline
\end{tabular*}

\end{table*}

The results presented in the previous sections are summarised in  \hyperref[tab:summary]{Table I}.
There are notable similarities and differences between the experiments, and in this section we will discuss these in detail.

\subsection{Similarities}
Both experiments produce long-lasting flows which advect magnetic fields towards the reconnection layer at around 50 km/s.
The flows collide at the mid-plane and form a dense reconnection layer which persists for over 200 ns, sufficient time for the development of complex features such as plasmoids.
The advected magnetic field is approximately constant in the plasma flows, and the magnetic flux is annihilated within a narrow layer.

In both cases, the layer width $\delta$ (measured using end-on interferometry and side-on Faraday rotation imaging) is slightly larger than, or roughly the same as, the ion skin-depth, $d_i=c/\omega_{pi}$, where $\omega_{pi}=Ze^2 n_e/\epsilon_0 m_e$ is the ion plasma frequency.
This is the condition for two-fluid effects to become important during reconnection,\cite{Yamada2006} although in our experiments the ion-ion mean free path ($\lambda_{i,i}$) is much smaller than any of the other length scales including $d_i$, which might result in collisions obscuring signatures of two-fluid effects.
Preliminary measurements of the out-of-plane magnetic field in a single experiment with carbon wires did not show a quadrupolar magnetic field, a classic signature of two fluid effects, and these measurements are the subject of further investigation.

For both experiments, the inflowing plasma is relatively cold, but undergoes substantial heating inside the reconnection layer, resulting in increased ionisation.
The ions are more strongly heated than the electrons, and the thermal energy gained by the ions is significantly larger than can be explained by the thermalisation of the flow kinetic energy, suggesting that the magnetic energy is converted into thermal energy.

The flows and magnetic field profiles are left/right symmetric, consistent with the expected even current division between the two arrays for both wire materials.
The initial conditions of the inflows have also been found to be reproducible between experiments, which justifies using time-resolved single-shot measurements across a range of shots to build up a global picture of the time evolution of the plasma conditions.

Side-on measurements of the reconnection layer in the ($x,z$) plane show that the flows and the reconnection layer are uniform in the $z$-direction perpendicular to the reconnection plane.
This enables line integrated measurements of electron density ($\int n_e(x,y,z) dz$) to be converted to average electron density by dividing by the out-of-plane length scale $h$, $n_e=\int n_e(x,y,z) dz/h$.
Additionally, it also suggests that the reconnection layer is not subject to any large scale instabilities with significant $k_z$ components, for example, due to current flowing in the layer.

\subsection{Differences}

\textsc{Inflows:} Although the inflow velocities ($V_{in}=50$ km/s) and thermal energy ($\bar{Z}T_e+T_i =110$ eV) were similar for the two materials, other flow parameters were different --- for example, the magnetic field was lower in aluminium (2 T) than carbon (3 T), and the electron density was higher (Al: \xcmcubed{5}{17}, C: \xcmcubed{3}{17}).
Combined with the large difference in ion mass (Al: 27 $m_p$, C: 12 $m_p$), this meant that the inflows were governed by significantly different dynamics in the aluminium and carbon cases.

For example, in aluminium the magnetic pressure was less than the ram pressure, giving a dynamic beta ($\beta_{dyn}=n_i m_i V^2/(B^2/2\mu_0)$) larger than one, $\beta_{dyn}\approx 10$.
This implies that the magnetic field is not dynamically significant in aluminium --- the field is still frozen into the flows ($Re_M=\mu_0 L V/\eta=27$), but it is advected as a passive quantity.
In contrast, for carbon $\beta_{dyn}=1.0$, which implies that the magnetic pressure and ram pressure are equally important, with the magnetic pressure playing a more significant role in the dynamics for carbon than for aluminium.

Similarly, we can compare the magnetic pressure to the thermal pressure, and calculate $\beta_{th}=n_i k_B(\bar{Z} T_e+T_i)/(B^2/2\mu_0)$ for the two cases.
As already noted, $\bar{Z} T_e+T_i$ is similar for aluminium and carbon, but the change in the other parameters gives $\beta_{th}=1.1$ for aluminium and $\beta_{th}=0.4$ for carbon.
Therefore the thermal pressure and magnetic pressures are similar in aluminium, implying both are much smaller than the ram pressure, placing these experiments in the strongly driven regime.
In contrast, for carbon the magnetic pressure is more than twice as large as the thermal pressure, suggesting that in the inflows the thermal pressure plays a lesser role compared to both the ram and magnetic pressures.

The flows from the aluminium wires are well-defined, super-sonic ($M_S=V/C_S=3.1$, where $C_S=\sqrt{(\bar{Z}T_e+T_i)/m_i}$ ) and super-Alfv\'enic ($M_A=V/V_A=2.2$, where $V_A=\sqrt{B^2/2\mu_0 m_i n_i}$), bounded by oblique shocks (similar to Ref. \onlinecite{Swadling2014}) and remain well-collimated until they reach the reconnection layer, presenting a highly modulated density profile.
Although the flows for carbon begin as distinct, well-defined streams,  they merge before reaching the reconnection layer, presenting a uniform inflow profile to to the layer. 
The flows in carbon are super-sonic ($M_S=1.7$), as in the aluminium case, but sub-Alfv\'enic ($M_A=0.7$), resulting in sub-fast-magnetosonic velocities ($M_{FMS}=0.65$).
This suggests a mechanism for smoothing out density gradients by fast-magnetosonic waves.

\textsc{Reconnection layer:} Clearly the long lasting nature of the layer suggests that there is an approximate balance between the total pressure in the inflows and the total pressure inside the layer for both carbon and aluminium.
The magnetic pressure at the mid-plane is close to zero as the magnetic field has all been reconnected, suggesting that the layer is supported by thermal pressure, which implies that compression, heating or flux pile-up at the layer occurs in order to achieve the overall pressure balance observed.
Heating is observed in both carbon and aluminium, as discussed below, but only in aluminium do we see evidence for flux pile-up, consistent with the super-Alfv\'enic flows.

For aluminium, the layer is relatively uniform despite the strong modulations in the inflows.
Although there are regions of enhanced density within the layer, these regions are symmetric about $y=0$ and slowly move outwards at the measured outflow speed.
However, in carbon  experiments, the layer is unstable despite the uniform inflow profile --- the layer rapidly breaks up into a chain of plasmoids which are visible in the electron density maps and the optical self-emission.
These plasmoids may play an important role in explaining the high electron and ion temperatures observed inside the reconnection layer.\cite{Hare2017, Hare2017c}

For aluminium, the layer density continually increases throughout the experiment.
The ion temperature within the reconnection layer decreases, with an abrupt change at \t{240}, which suggests a complex physical process, which we discuss in more detail below.
In contrast, in carbon experiments the electron density inside the layer plateaus after \t{230} (Ref. \onlinecite{Hare2017c}, Fig. 5a), and there is no sudden change in layer properties.
Therefore, perhaps counter-intuitively, although the layer appears far more unstable in carbon than in aluminium, for carbon the average plasma parameters do not change significantly as the layer evolves in time whereas there are significant changes in aluminium.

Further evidence for a sudden change in the layer parameters in aluminium comes from Thomson scattering.
Ion acoustic peaks are observed in the the Thomson scattering spectrum from plasma inside the layer (\sref{fig:al_thomson}{b}) at \t{210}, and these peaks become more pronounced after \t{240} (\sref{fig:al_thomson}{c}).
This implies a sudden decrease in the ion temperature (from $T_i=270\rightarrow30$ eV), as ion thermal motion is responsible for broadening the ion-acoustic peaks.
For aluminium, the electron temperature only increases slightly as the plasma enters the reconnection layer, increasing from 15 eV to 40 eV.

In carbon, the ion acoustic peaks are not sufficiently separated, resulting in a flat-top spectrum (\sref{fig:c_thomson}{d}) which implies that $\bar{Z}T_e\approx T_i$, the condition for broadening by ion-thermal motion to make the two ion-acoustic peaks indistinguishable.
In these experiments the electrons are significantly heated by a factor of around six as the plasma enters the reconnection layer ($T_e=15\rightarrow 100$ eV).
The ions are heated even more significantly ($T_e=50\rightarrow 600$ eV), but the measured ion temperature does not change significantly inside the layer as it evolves over a 200 ns timescale.

These differences are partially explained by the differences in the radiative cooling time-scale --- for aluminium this is relatively short, $\tau_{rad}\approx4$ ns,\cite{Suzuki-Vidal2015} whereas for carbon the electrons are not significantly cooled on the experimental time-scale, with $\tau_{rad}\approx 600$ ns.\cite{Hare2017c}
The electron-ion energy exchange time is also quite short for aluminium, $\tau_{E, ei}\approx 40$ ns, dropping to $\tau_{E, ei}\approx 20$ ns as the electron density increases.
Hence for aluminium, the electrons can be an efficient sink of energy for the ions, whereas in carbon this time-scale is always longer than the experimental timescale, $\tau_{E,ei}\approx 600$ ns.

The ion-electron equilibration time-scale decreases as the electron density increases, which could account for the correlation between the sudden drop in ion temperature as the  electron density rises in experiments with aluminium.
Although the exact mechanism to link these two events is unclear, any mechanism which increases electron density by reducing the ion temperature would clearly be unstable, and should result in cold ions inside a dense layer as we observe.

The measured plasma parameters in \hyperref[tab:summary]{Table I} can be used to calculate another important dimensionless quantity for magnetic reconnection, the Lundquist number $S=\mu_0 V_A L/\eta$.
In aluminium, the Lundquist number is modest, around 10, but it is over an order of magnitude higher in carbon at 120, which is primarily due to the higher electron temperature in carbon resulting in a decrease to the plasma resistivity as $\eta\propto T_e^{-3/2}$.
The criteria for plasmoid formation depends on the Lundquist number, as well as the ratio of the layer half-length to the ion skin-depth ($L/d_i$) for plasmoids in the semi-collisional regime.
For the parameters measured, plasmoids are expected to form in carbon, but not in aluminium, consistent with our observations.\cite{Loureiro2015,Hare2017c}

There are clear differences between experiments with aluminium and carbon wires.
The inflows in aluminium are more strongly driven and denser than those in carbon, and with more pronounced density modulations.
The reconnection layer is also significantly different in the two cases, with a uniform layer inside which the parameters continuously evolve for aluminium, and an unstable layer which remains unstable over the duration of the experiment for carbon.

In the inflows, the dynamics and density modulations can be attributed to the dimensionless plasma parameters, such as the sonic and Alfv\'enic Mach numbers.
Inside the layer, species-dependent effects such as radiative cooling and ion-electron equilibration appear to play a more important role.
The ability to access these very different regimes using the same physical hardware allows us to study reconnection over a wide range of parameter space using the same suite of diagnostics, and thus perform detailed quantitative comparisons between plasma regimes in which many of the variables are held constant.

\section{Results from Varying the Array Geometry}\label{sec:geometry}

\begin{figure*}[!h]
	\Fig{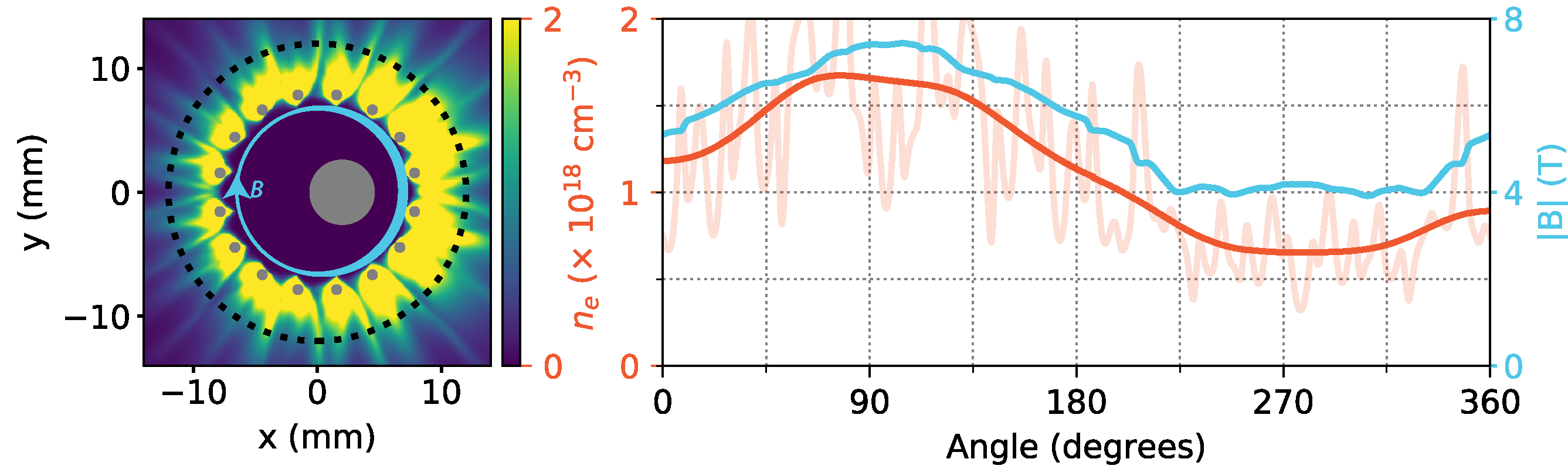}
	\centering
	\caption{Simulation data from the resistive MHD code, Gorgon. a) Line averaged electron density ($\int n_e dz/h$) map from an exploding wire array with a displaced thick electrode (large grey circle). The azimuthal variation of the driving magnetic field at the wires is sketched in blue. The profiles in b) are taken along a circle at 12 mm radius, indicated by a dashed black line, with the profiles starting at ($x,y$)=(0,12) mm and proceeding clockwise. b) Profiles of $n_e$ (orange) and $|B|$ (blue). For $n_e$ we show both the raw data from the simulation, which is highly modulated by shocks, and a smoothed version to guide the eye.}
\end{figure*}

\begin{figure*}[t]
	\Fig{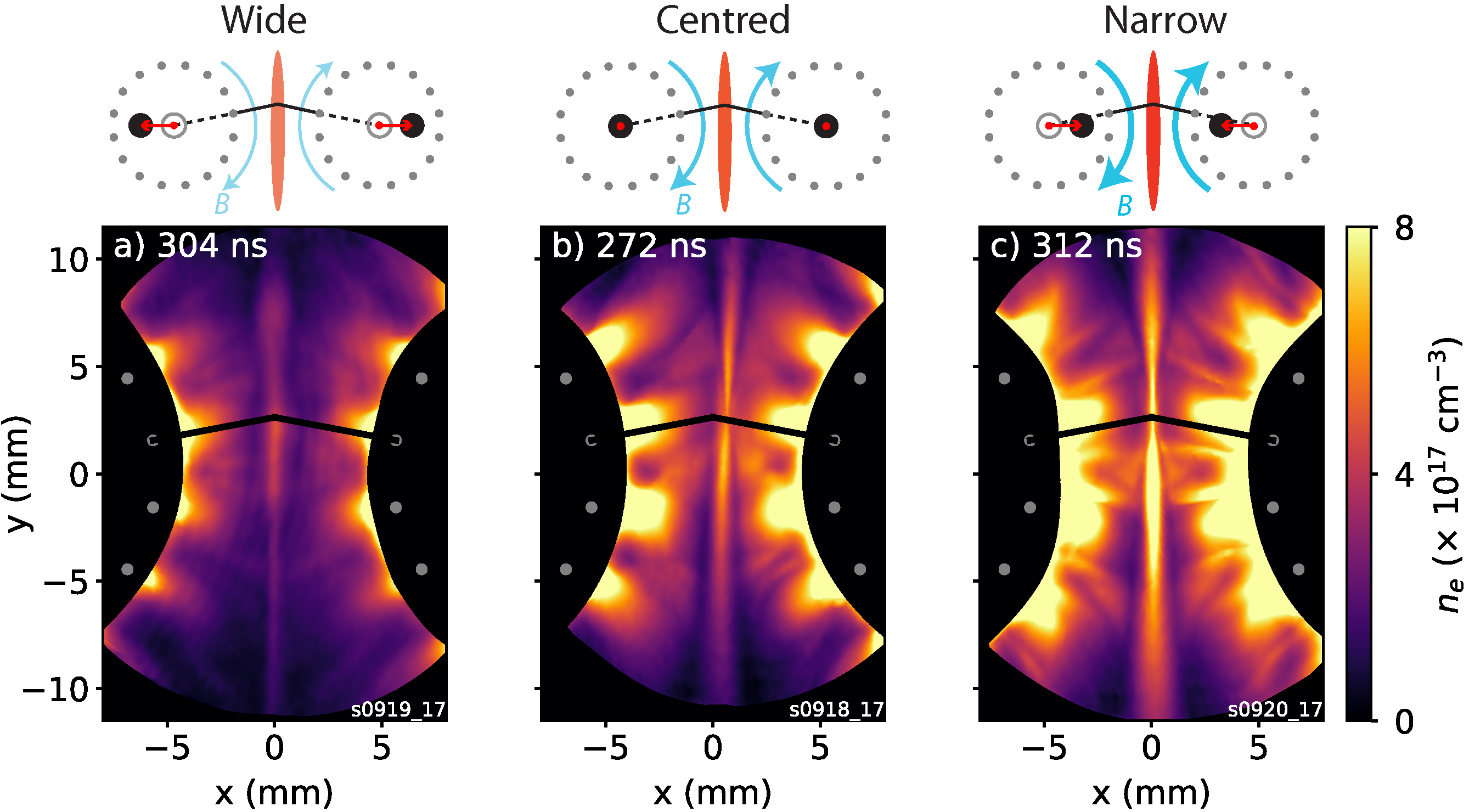}
	\centering
	\caption{Three different electrode configurations to produce varying density in the inflowing plasma. The grey outlines in the upper graphics show the standard (`centred') position of the thick electrode, with a red arrow showing the displacement for the wide and narrow case (exaggerated scale for clarity). Electron density maps are shown below using the same colour map, with wire positions marked with solid grey circles. Dashed black lines mark a chord joining the centre of the arrays to a wire position, which is extrapolated as a solid black line to the mid-plane. a) Wide configuration, with the central electrodes displaced 2 mm away from the centre of the two exploding wire arrays, further away from the reconnection layer. b) Centred configuration, with the electrodes at the centres of the two exploding wire array. c) Narrow configuration, electrodes displaced 2 mm away from the centre, towards the reconnection layer.}
\end{figure*}

In addition to varying the wire material, we can also significantly affect the plasma parameters in the inflows and the reconnection layer by varying the geometry of the exploding wire arrays.

This new geometry is overlaid onto the simulation data in \sref{fig:simulation_lineouts}{a}, in which we displace the thick electrode inside the wire array.
This results in an azimuthal variation in the strength of the driving $\textbf{J}\times\textbf{B}$ force felt by the wires, with a stronger driving force on those wires closest to the electrode and a weaker for those wires further away.
This is due to two factors: a change in the magnetic field from the thick electrode at the wires, as well as a change in the division of current between the wires.

Firstly, the driving magnetic field ($\textbf{B}$) from the thick electrode falls off roughly as $1/r$, and so as the distance between the thick electrode and those wires decreases (increases), then $\textbf{B}$ increases (decreases).
Secondly, wires further from the thick electrode now represent a higher-inductance path than the wires closer to the electrode, as the current path through the more distant wires encloses a larger area.\cite{Velikovich2002}
The decreased (increased) inductance increases (reduces) the fraction of the current ($\textbf{J}$) flowing through the wires closer to (further from) the electrode.
The net result is therefore a higher (lower) $\textbf{J}\times\textbf{B}$ force on wires closer to (further from) the thick electrode.
This should lead to larger densities and advected magnetic fields in the flows from the wires closer to the electrode than in the flows from wires further from the electrode.

We study the effect of this change in driving force on the ablation flows from the wires using the Gorgon code to run three-dimensional resistive MHD simulations (\sref{fig:simulation_lineouts}).\cite{Chittenden2004,Ciardi2007}
A line averaged electron density map ($\int n_e(x,y,z) dz/h$) is shown in \sref{fig:simulation_lineouts}{a}, where there is a clear azimuthal variation.
Results are shown in \sref{fig:simulation_lineouts}{b} as profiles along a circle with a radius of 12 mm --- the profiles start at ($x,y$)=(0,12) mm and proceed clockwise.
The electron density $n_e$ is shown in orange, with both the raw data from the simulation (lighter orange) and a smoothed profile (dark orange).
The profile is highly modulated due to the presence of multiple shocks per wire and so we provide a smoothed version to show the overall trend.
From the smoothed data we can see a variation in $n_e$ greater than a factor of 2, from around \xcmcubed{0.6-1.7}{18}.
The magnetic field magnitude also increases in the flows from the wires closest to the electrode, varying from $B=3.9-7.5$ T, less than a factor of two.

These simulations suggest we can significantly affect the parameters of the ablation flows coming from an exploding wire array, and we can now use this new geometry in our reconnection experiments by placing two such arrays side by side and driving them in parallel.
Three such geometries are demonstrated in \sref{fig:inflow_density_scan}{}.
The standard, or `centred' case is shown in \sref{fig:inflow_density_scan}{b}, where the two thick electrodes are at the centre of the wire arrays.
This was the configuration used in all of the results presented previously in this paper.

We can then move the two thick electrodes 2 mm further from, or closer to, the mid-plane between the two arrays --- see \sref{fig:inflow_density_scan}{a} for the `wide' case and \sref{fig:inflow_density_scan}{c} for the `narrow' case.
This has a significant effect on the nature of the plasma inflows and on the reconnection layer --- for the `wide' case the flows are less dense and carry a smaller magnetic field than the centred case, but in the `narrow' case the flows should be denser and carry a larger magnetic field.

Three electron density maps are shown in \sref{fig:inflow_density_scan}{}, corresponding to the three electrode geometries described above.
The experiments were carried out using carbon wires, and the arrays were otherwise identical to those described in the previous section.
The three maps in \sref{fig:inflow_density_scan}{} are shown using the same colour scale, and it is clear that there is a significant increase in density, both in the flows and inside the reconnection layer, as the electrodes move from the wide configuration to the narrow configuration.
Although these electron density maps were taken at slightly different times during the current pulse, we know that the electron density inside the layer does not change significantly for carbon after around \t{230} (see Ref. \onlinecite{Hare2017c}, Fig. 5a), and so it is reasonable to compare layer electron densities in the layer between the three cases.
In the wide case, the electron density inside the layer is around $n_e=$ \xcmcubed{4}{17}, which rises to $n_e=$  \xcmcubed{6}{17} in the centred case and further to $n_e=$  \xcmcubed{11}{17} in the narrow case.
This is an increase of around a factor of three, suggesting we can access significantly different densities using this experimental configuration.

Another interesting change is the divergence of the outflows from the two wire arrays.
In a simple model, the flow direction is set by the $\textbf{J}\times\textbf{B}$ force.
For the centred case, this is simply radially outwards, but for the wide and narrow cases the magnetic field is no longer purely tangential to the circle on which the wires lie, and so the flow direction should differ.

In \sref{fig:inflow_density_scan}{}, black solid lines are drawn on both the diagrams above and the electron density maps below, representing a chord which joins the centre of the arrays to one wire, and extrapolates this chord out to the mid-plane.
We can see that for the centred case the flows do not follow these lines, and instead appear to diverge more than predicted by this simple mode.
This suggests that there are additional forces acting on the flows --- not only do the flows carry current (as they must do in order to advect magnetic field) but the reconnection layer also carries current, and these currents create a more complex magnetic field topology which acts on the flows.
For the wide case, the flows are closer to the lines, with the angle between the flows smaller than in the centred case.
For the narrow case the flows are further from the lines, at a larger angle than in the centred case, which is primarily due to the overall change in topology caused by moving the thick electrode closer to the mid-plane.

The divergence of the flows is important because it affects how we calculate the length scale $L$ of the reconnection layer, an important quantity which is used in many calculations, such as the Lundquist number $S=\mu_0 L V_A/\eta$ and the plasmoid instability (in the form of $L/d_i$).
We have previously used $L=R_C/2$, where $R_C$ is the radius of curvature of the magnetic field lines at the mid-plane, which we determined geometrically as half the separation of the thick electrodes.
Therefore, by moving the electrodes further apart we expect the radius of curvature and hence the length scale of the reconnection layer to increase, and qualitatively this is confirmed by our observations of the angle between the inflows propagating towards the mid-plane.

Preliminary experiments with Faraday Rotation Imaging have confirmed that the magnetic field advected by the flows is also affected by the separation between the thick electrodes.
This is expected because the flows advect a fraction of the driving magnetic field, and, as we previously discussed, this driving magnetic field is weaker in the wide case and stronger in the narrow case due to the distance between the thick electrode and the wires.

We have not yet used Thomson Scattering to measure the flow velocity and the electron and ion temperatures for these new configurations.
This would enable us to calculate the sonic and Alfv\'enic Mach numbers of the flows.
The formation of oblique shocks between flows from adjacent wires has been observed in the end-on electron density map for the narrow case \sref{fig:inflow_density_scan}{c}, which suggests that the flows are super-sonic and super-Alfv\'enic for this case.
If so, this experimental configuration may be another way to reach the strongly driven regime of reconnection, but without the radiative cooling inherent in the experiments with aluminium wires.

So far we have demonstrated three significant changes to the plasma parameters by changing the electrode separation --- the density of the inflows, the magnetic field advected by the inflows, and the length scale of the reconnection layer, set by the radius of curvature of the reconnecting magnetic field lines at the mid-plane.
In the previous section we also demonstrated how to change the parameters of the inflows by changing the wire material, but this introduced further variation due to species-specific effects such as radiative cooling.
By keeping the wire material fixed and changing the separation of the electrodes we gain additional control over the plasma parameters.

We hope to use this platform to study the regions of parameter space in which plasmoids form.
Our previous work with the centred case showed that the plasma was consistent with the semi-collisional regime of the plasmoid instability,\cite{Hare2017c} given by $(L/d_i)^{8/5}\ll S \ll (L/d_i)^2$, where $d_i=c/\omega_{pi}$, the ion skin-depth, and $\omega_{pi}$ is the ion plasma frequency.
These boundaries were obtained using an asymptotic theory in which the relevant length scales --- such as the ion and electron skin depths, the collisional mean-free-path and the width of the layer --- all differed by many orders of magnitude.

This is not the case in our experiments, and indeed it is difficult to realise this scale separation in any experimental or computational system.
However, the fact that the theory has only been established in the asymptotic limit does not imply that plasmoids cannot form in smaller scale systems as well.
It would be useful to conduct experiments with a platform capable of varying the plasma parameters in a precise, measurable and controllable manner, to enable the study of the stability boundaries for the formation of plasmoids in realistic physical systems.
The results presented in this section offer a first step on one path to realising such an experimental platform.

\section{Asymmetric reconnection}\label{sec:asymmetric}

\begin{figure}[t]
	\Fig{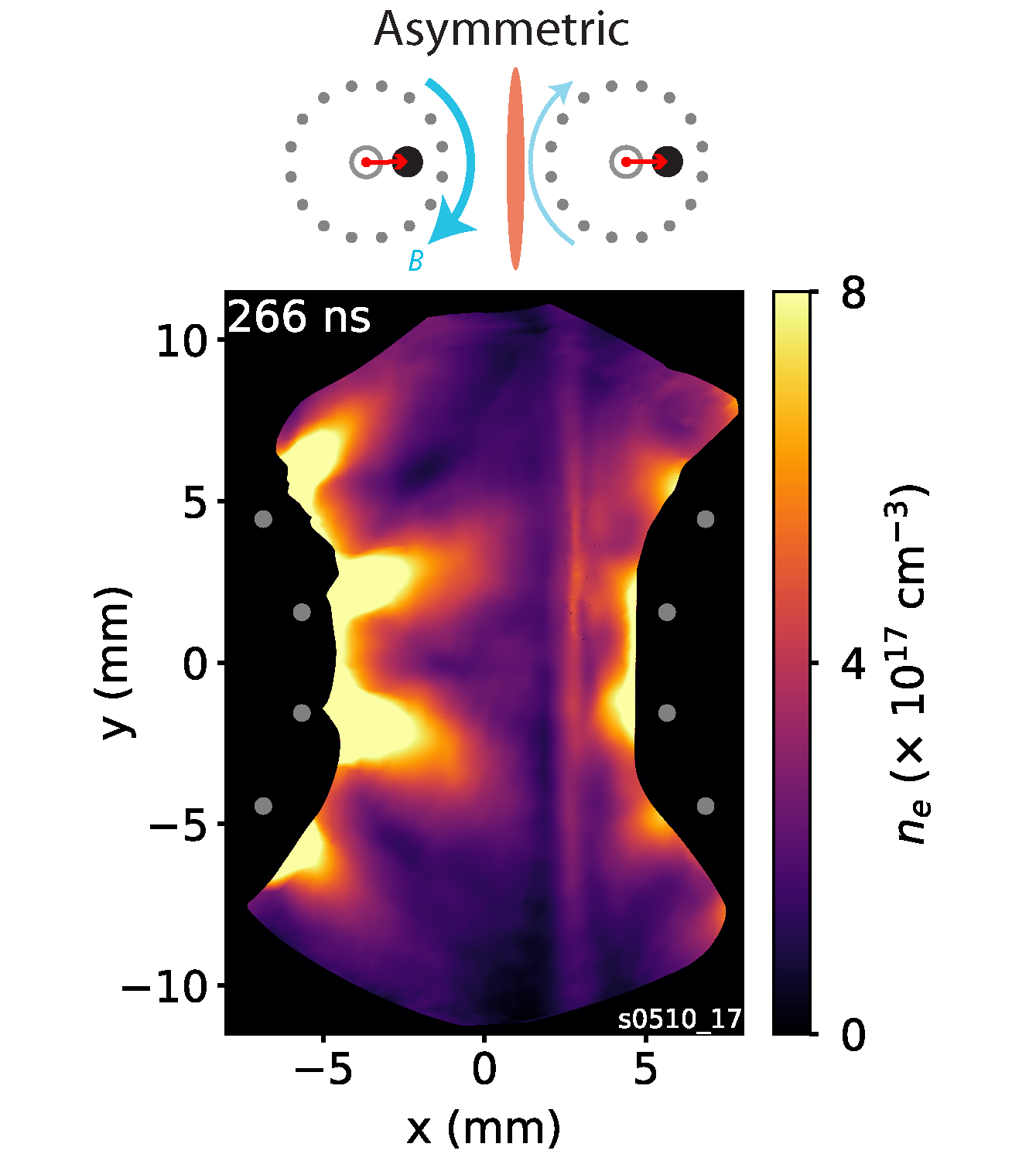}
	\centering
	\caption{An electrode configuration to produce asymmetric reconnection. One electrode is displaced closer to the mid-plane between the two arrays, and one electrode is displaced further away. The position of the wires is marked with grey circles. The reconnection layer has been displaced in the same direction as the electrodes, to the right, a result of the asymmetric drive of the flows from the left and right wire arrays.}
\end{figure}

In the previous section we presented a technique for varying the parameters in the inflows by altering the driving force through changing the separation of the thick electrodes inside the two wire arrays.
This technique naturally suggests another configuration for this experimental platform, in which we can study asymmetric reconnection, and this configuration is shown in \sref{fig:asymmetric}{}.

Here, the two thick electrodes inside the wire arrays are displaced in the same direction from their central location, by 2 mm to the right.
This changes the magnetic field strength at the wires closest to the mid-plane, as discussed in the previous section.
For the left-hand array, these wires experience a larger magnetic field, as the driving magnetic field produced by the thick electrode falls off roughly as $1/r$.
These wires also carry a larger fraction of the current as they represent a lower inductance path, and this, combined with the increased magnetic field, results in denser flows which are expected to advect a larger magnetic field.
For the right-hand array, the opposite argument applies, with the wires being driven by a smaller magnetic field and carrying a lower current, and therefore producing less dense flows advecting smaller magnetic fields.

Hence when the plasma flows collide, they are mis-matched. 
For this preliminary work we do not have an image capturing the moment of collision, but we know that the layer has formed by \t{206}, measured using a gated XUV pinhole camera.
At this time, the layer already appears to be displaced far to the right, suggesting two possibilities: 
firstly, that the flow velocity of the right hand array is lower than the left hand, and so the point of collision is closer to the right array, or secondly that the plasma from the left array quickly sweeps up the plasma from the right array, and it is this piled up plasma we see displaced to the right.

The asymmetric nature of the inflows and layer can clearly be seen in the electron density map in \sref{fig:asymmetric}{}.
On the left, the inflows propagate with a large angle between them, as expected due to the position of the thick electrode close to the mid-plane.
On the right, the inflows do not extend far enough into the field of view for the angle between them to be measurable, and the reconnection layer is at \x{2.5}.
The layer is denser than the plasma flowing in from the left or the right, and there is a pronounced depletion of electron density on the left-hand side.

The layer position is roughly constant in time --- a second frame of interferometry, taken during the same shot, 20 ns later (at \t{286}) showed that the layer had not moved appreciably within the limit of the diagnostic resolution.
This suggests that the layer has reached an approximate pressure balance, where the flows from the left-hand array have diverged sufficiently that the total pressure they exert on the layer is matched by the pressure exerted on the layer by the flows from the right-hand array.
This is convenient, because in order to make detailed measurements with Thomson Scattering, we need an object which remains at a well defined position so that the probe laser can be aligned through the layer (in a geometry similar to \sref{fig:c_thomson}{a}).

We have not yet used Faraday Rotation Imaging to diagnose the asymmetry in the reconnecting magnetic fields, or the location of the current sheet.
In asymmetric reconnection the current sheet (where the magnetic field changes sign) and the stagnation point (where the plasma density piles up) are not at the same location, being displaced due to the asymmetric drive on the reconnection layer\cite{Cassak2007, Yoo2014,Yoo2017}.
It would be interesting to use this platform to study this phenomena in more detail, as it is important for understanding in-situ measurements of asymmetric reconnection by satellites in the Earth's magnetosphere.

As with the configurations discussed above, we can scan a wide range of inflow parameters, and hence asymmetries, by changing the position of the electrodes.
This may make it possible to reach a regime in which plasmoids form during asymmetric reconnection, which we do not believe has previously been demonstrated in the laboratory, but may occur in the Earth's magnetosphere.

\section{Conclusions}\label{sec:conclusions}

In this paper we have provided a detailed description of the versatile pulsed-power driven magnetic reconnection platform developed on the \textsc{Magpie} generator at Imperial College London.
This platform uses two exploding wire arrays, in which initially solid wires are converted to plasma by a current pulse and accelerated radially outwards by a large $\textbf{J}\times\textbf{B}$ force.
The plasma flows collide at the mid-plane between the two arrays, and magnetic reconnection occurs as oppositely directed magnetic fields, advected by the two flows, annihilate inside an intense sheet of electric current ($J\approx 3$ GA/m$^{2}$).

We first demonstrated the versatility of this platform through experimental results using either aluminium or carbon wires, in which we showed that the plasma parameters in the inflows, such as the thermal and dynamic betas, differed significantly between the two materials.
In aluminium, the ram pressure dominated over the magnetic and thermal pressures, which were of the same order, giving super-sonic and super-Alfv\'enic inflows with large density modulations caused by the formation of oblique shocks between the outflows from the discrete wires.
In carbon, the magnetic and ram pressures were of the same order, and both were slightly larger than the thermal pressure, resulting in super-sonic but sub-Alfv\'enic flows which merged as they propagated outwards, presenting a smooth density distribution to the reconnection layer. 

These inflows collided at the mid-plane to produce a well-defined reconnection layer, which possessed markedly different properties depending on the material of the original wires.
In aluminium, the ions were initially hot inside the reconnection layer, before rapidly cooling around \t{240} after current start.
Magnetic flux was observed to pile up outside the reconnection layer, consistent with the super-Alfv\'enic nature of the inflows.
For carbon, the electrons and ions were heated significantly, and remained hot over the entire life-time of the layer.
Inside the layer, the average electron density did not change significantly over 200 ns, but the layer became highly unstable, forming plasmoids which were ejected out of the layer at super-Alfv\'enic velocities.

We then described a second technique for varying the plasma parameters, in which the thick electrodes within the wire arrays were displaced to change the driving force on the wires closest to the mid-plane.
This technique allowed us to change both the density and the magnetic field in the inflows, resulting in different densities and currents inside the reconnection layer, and opening the way to studying the formation of plasmoids in a wide range of plasma parameters.

Finally, we discussed an extension of the second technique, in which the left and right arrays were driven with different strengths, giving asymmetric reconnection.
We observed the formation of a reconnection layer which was significantly displaced from the mid-plane between the arrays, suggesting that the flows were indeed asymmetric.
After formation, the layer was not observed to move further with time, which should make it possible to study the layer in more detail using Thomson scattering.

These three related experimental set-ups show the breadth of plasma physics we can access in a precise, reproducible and controlled manner, and will allow future experiments to study effects such as the importance of radiative cooling, ionisation, plasmoid formation and the displacement of the stagnation layer.

\section*{Acknowledgements}

This work was supported in part by the Engineering and Physical Sciences Research Council (EPSRC) Grant No. EP/N013379/1, and by the U.S. Department of Energy (DOE) Awards No. DE-F03-02NA00057 and No. DE-SC-0001063. AC and NFL were supported by LABEX Plas@Par with French state funds managed by the ANR within the Investissements d'Avenir programme under reference ANR-11-IDEX-0004-02. NFL was supported by the NSF-DOE partnership in basic plasma science and engineering, award no. DE-SC0016215

\bibliography{library}

\end{document}